%% file: UnityOS_Vision-arXiv.tex
\documentclass[sigconf]{acmart}
\renewcommand\footnotetextcopyrightpermission[1]{} % removes footnote with conference information in first column
\usepackage{booktabs} % For formal tables

\begin{document}
\title{The Case for a Single System Image for Personal Devices}
%\titlenote{Produces the permission block, and
%  copyright information}
%\subtitle{Extended Abstract}
%\subtitlenote{The full version of the author's guide is available as
%  \texttt{acmart.pdf} document}

\author{
{\rm Beom Heyn Kim, Eyal de Lara, David Lie}\\
University of Toronto}

\begin{abstract}
Computing technology has gotten cheaper and more powerful, allowing users to have a growing number of personal computing devices at their disposal. While this trend is beneficial for the user, it also creates a growing management burden for the user. Each device must be managed independently and users must repeat the same management tasks on the each device, such as updating software, changing configurations, backup, and replicating data for availability. To prevent the management burden from increasing with the number of devices, we propose that all devices run a single system image called a personal computing image. Personal
computing images export a device-specific user interface on each device, but provide a consistent view of application and operating state across all devices. As a result, management tasks can be performed once on any device and will be automatically propagated to all other devices belonging to the user. We discuss evolutionary steps that can be taken to achieve personal computing images for devices and elaborate on challenges that we believe building such systems will face.
\end{abstract}

\maketitle

\input{intro}
\input{alternatives}
\input{usage}

\input{evolution}

\input{challenges}
\input{conclusion}

\bibliographystyle{ACM-Reference-Format}
\bibliography{bk} 

\end{document}

%% file: intro.tex
\section{Introduction}\label{sec:intro}

Improvements in technology, increases in production volumes and market competition have created an environment where the average user may easily own several personal computing devices. For example, a user may typically have one or more desktops or laptops, a smartphone, a tablet, a game console, as well as various media and storage appliances. Each of these devices has an abundance of storage and computational power, with latest smartphones sporting processors with clock speeds of more than 1GHz and several cores. The power and storage capabilities of these devices enable them to run full-fledged operating systems (OS) that support rich, feature-packed third-party applications, just like their more traditional non-mobile desktop counterparts. This trend towards a greater number of personal computing devices per user is likely to continue as the computational power that can be purchased for a given price increases, and as device manufacturers push to introduce extensibility into other traditionally static devices, such as automobiles, home climate control and home appliances.

While the ability to own a variety of personal computing devices is largely beneficial for the average user, it has also created a growing maintenance burden for the user. This burden manifests itself in at least two ways. First, the user's data and applications, instead of being contained in a single device, are now distributed across multiple devices. While solutions exist to replicate data, such as DropBox and LiveMesh, these do not replicate applications -- users must manually install the applications they want on all devices and keep them in sync by replicating configuration changes, patches and updates. The management of applications can impose a non-negligible burden on the user of these devices. Consider a user with 5 computing devices and about 20 applications that are common between all devices. If there was only one management task for each of those applications per year, this would still result in 100 management tasks per year or roughly one change action every 4 days! Second, the user incurs a heavy management cost whenever she acquires a new device or replaces a failed one. She must configure the device, for example, by setting network parameters and account credentials. In addition, she must install all her applications and copy over any data that she wants to access with those applications. Current computing trends point to even more powerful devices, with more capabilities and more storage available at ever decreasing costs in the future. As the number of devices increases, so will this ``tyranny of devices" -- as users will be forced to either spend more and more time managing their devices, or simply stop acquiring new devices.

This increased management load stems from the fact that while all of the user's devices are able to communicate with each other, they operate as separate system images, isolated from each other. We believe that having multiple personal computing devices that are separate islands of computing goes against the computing model that users desire. Instead, users want all devices to be identical except for their location and form factor. Users want a smartphone with a small screen to fit in their pocket and they want a desktop with a large screen when they are in their office or at home. However, anything that does not have to do with the physical properties of a device should be identical across all devices. Users want to have the same applications, same data and same configuration on all devices, and they want all those things to be kept synchronized across devices without their intervention. In essence, we believe that users want a single system image for all their personal computing devices. This personal computing image would give all devices a unified object name space, where an object could be a piece of data or an application -- any change to an object is immediately made visible to other devices. In this way, any management tasks would only have to be performed once on any device and would be automatically propagated to all of the user's other devices.

The idea of distributing computation over many devices is not new -- a variety of proposals in mobile and pervasive computing have made various cases for distributed computing over devices. Moreover, the illusion of a single system personal computing image can likely be achieved in some capacity with a variety of well known mechanisms, such as replication, remote terminals and process migration. However, we believe the key difference between our proposal and previous proposals lies in the assumptions and the goal of the distributed computation. Previous systems assumed weak devices, and the goal of distributing computation was to improve performance, to increase storage capabilities or to save energy. We believe that in the cases where the underlying hardware has not already exceeded the requirements for adequate performance, storage or battery life for our devices, they will eventually. One only has to consider recent technology trends. For example, a typical tablet or smartphone in 2011 had a dual core ARM processor clocked at 1.5 GHz and 1 GB of RAM. In 2007, which was merely 4 years earlier, the average personal computer had a dual core Pentium processor clocked at 1.8GHz and 1-2GB of RAM! Nowadays, smartphones are more powerful than the average personal computer from 2007. Since performance benefits of distributing an application over several devices will only exist for 2-3 years before the underlying technology obsoletes it, we believe performance alone is not enough for the reason to make distributed computing broadly applicable on devices. Going forward, we believe the primary reason to distribute computation is to reduce management effort as this cost will not go away no matter how much the underlying hardware improves. Instead, this cost will only increase as the number of personal computing devices a person has continues to increase. The assumptions about the availability of resources in the underlying hardware naturally restrict personal computing images to devices that can support a general purpose OS with 3rd party applications. Simpler devices with limited functionality, such as wireless sensors or bluetooth headsets, may still be accessible to devices running the personal computing image, but do not incur the same sort of management burden and would not run the personal computing image themselves.

In this paper, we begin in Section~\ref{sec:alternatives} by elaborating on some alternative solutions that have been proposed or exist currently, and argue why none of them can implement a personal computing image adequately. Then, in Section~\ref{sec:usage}, we discuss the usage model for our vision of personal computing images. In Section~\ref{sec:evolution}, we give our view of how current systems can evolve towards personal computing images and introduce a system called UnityOS, which we are building to realize personal computing images. Finally, we discuss challenges to implementing UnityOS in Section~\ref{sec:challenges} and conclude the paper in Section~\ref{sec:conclusion}.

%% file: alternatives.tex
\section{Alternative Solutions}\label{sec:alternatives}

In this section, we review some of the previous and current proposals for supporting the growth in personal devices and elaborate on how our vision of a personal computing image differs from them.

\paragraph{\textbf{Pervasive Computing:}} Our proposal has many similarities with the goals of pervasive computing, which aims to have computing devices, embedded in a wide variety of objects around us, work together to enhance our lives. While pervasive computing certainly has the concept of a seamless computing experience across devices, this is not necessarily done with a single system image. For example, systems like One.world~\cite{one.world:Grimm:2004:SSP:1035582.1035584} and Gaia~\cite{Roman02gaia:a}, provide frameworks in which an application may run in a distributed manner across devices. Their goal is to enable applications to follow users around and merge into the background. In a personal computing image, it is the OS that must operate in a distributed manner, not the applications. Our goal of simplifying device management is orthogonal to whether applications are distributed or not. An application may only be able to run on a single device at a time, or it may be capable of using several devices simultaneously, so long as the underlying OS ensures that changes and updates made on one device are consistently and instantly visible to other devices.

\paragraph{\textbf{Distributed Computation and Offloading:}} Several recent proposals~\cite{clonecloud:conf/eurosys/ChunIMNP11, maui:Cuervo:2010:MMS:1814433.1814441} aim to offload computation from mobile devices onto fixed servers running in the cloud. The goal of this offloading is two-fold: to shorten execution time and to reduce battery usage on the mobile device. Similarly, other work looks to improve the performance of applications by distributing computation over heterogeneous processors so that each distributed component is best able to take advantage of the hardware it is running on~\cite{barrelfish:multikernel:Baumann:2009:MNO:1629575.1629579, helios:Nightingale:2009:HHM:1629575.1629597}. While offloading and distributed applications may also improve the performance of applications on a personal computing image OS, performance is not the primary goal. A personal computing image's goal is to ensure that all devices appear to be running the same OS image, with the same applications and data so that management tasks do not need to be repeated on each device. Thus, while personal computing images may borrow many techniques from distributed computation, such as weak consistency, process migration and replication, its primary goals are reducing user maintenance effort and increasing the availability of data and applications to the user.

\paragraph{\textbf{Web Applications:}} Web applications such as Google Docs, Office Web Apps and Flicker take what would normally be an application running locally on a device and implement them as a distributed service over the Web. Because all persistent state is stored on the server, users perceive consistent application state and data across all devices that they access the web application with. While these are exactly the same goals as personal computing images, web applications only solve the problem for one application at a time. Turning a traditional desktop application into a web application is labor intensive process -- not only does the code need to be rewritten, but also application developers must carefully and manually plan how state is to be distributed and transmitted back and forth between the user's device and the application server. In contrast, by implementing the distribution at the OS layer, personal computing images achieve consistency for all applications regardless of whether the applications are distributed or not.

\paragraph{\textbf{Distributed Operating Systems:}} Personal computing images actually have the most in common with both shared and non-shared memory distributed OSs~\cite{hive:Chapin:1995:HFC:224056.224059, k42:Silva:2006:KIO:1131322.1131333, tornado:Gamsa:1999:TML:296806.296814, cellulardisco:Govil:1999:CDR:319344.319162, amoeba, sprite}. While early systems have separate address spaces for each node and later systems have hardware-supported shared-memory, they all share a common characteristic -- they implement a single system image over a network of tightly coupled compute nodes. Each node has its own processing and memory resources, but the OS abstracted this away to make them appear as one large pool of resources that is administered as a single entity. A consistent view of file storage and the state of other applications is provided to each program on the system. The main difference between the personal computing image and these distributed OSs is in the assumptions they make on the underlying hardware. These systems assume fast (for the time), highly reliable networks between a homogeneous set of nodes and in most cases, hardware support for memory consistency. The exorbitant cost of such hardware forced the designers to also assume that the systems would be heavily utilized, possibly by more than one user, so there would be large amounts of concurrency at both the application and OS layers. Since all the devices in a personal computing image are owned by a single person, most of the devices will be idle most of the time and there will be relatively little concurrency. As a result, there will also be less of a reliance on fast reliable networks, and some devices might have only a slow, low-bandwidth, high-latency network connection to other devices. Finally, a user's devices will necessarily be heterogeneous in terms of resources and computing power. 

In summary, we believe the salient properties that make personal computing images different from previous work and other solutions are as follows:
\begin{enumerate}
\item The primary goal of a personal computing image is to create the illusion to the user that all their devices are running a single, consistent system image. In this way, management tasks performed on one device are immediately propagated to and visible on other devices.
\item We aim to make the OS distributed, not the applications that run on the OS. Even if an application is not distributed and can only execute on a single node, the OS will ensure that any other device that interacts with the application or its data sees the most recent version of that state.
\item Devices are heterogeneous in their computing and storage capabilities, and may have varying qualities of network connection to each other. However, personal devices and applications are normally used by a single user, so they will tend to be lightly loaded and while users may switch between devices frequently there will be little true concurrency among the devices.
\end{enumerate}

%% file: usage.tex
\section{Usage Model}\label{sec:usage}

Different assumptions about the computing hardware and its availability mean that personal devices will have a different usage model than other computing devices have had in the past. Never before have computing devices been so cheap and so powerful that a single user may own more than a handful of them. However, we believe that users will own dozens of personal computing devices capable of running third party applications and performing a wide range of tasks, all reserved for the exclusive use of a single user. Today, personal computing devices consist of items such as our desktop and laptop personal computers, smartphones and tablets. In the future, these will extend to our other possessions such as our homes, cars, televisions and household appliances.

The usage model of personal computing images will be similar in some ways to recent data replication solutions, such as DropBox and OneDrive, which automatically replicate data across user devices. They will perform eager replication with the assumption that concurrent access and conflicts are rare, and in the absence of conflicts, do so entirely without the manual intervention of the user. We have the same goal for personal computing images. They should propagate OS state, application state and user data so as to give all devices a consistent view of that state without the intervention of the user so long as there are no conflicts. Because of the limited amount of concurrency and reasonable ease at which devices can gain some level of network connectivity, we assume conflicts will be rare, but when they do happen, they will require manual resolution by the user.

However, unlike current data replication systems, personal computing images function at the OS level so they will be aware of the heterogeneity of the user's devices, whereas current replication solutions are agnostic to it. Users will naturally not access every piece of data they have from every device they own -- there will be some devices that will never access some data. There are various reasons for this. For example, data objects will not be accessed from devices that are unable to manipulate them -- a device with a low resolution display or low computing power need not have high quality video files replicated to it. In some ways, this shares the same philosophy as the EnsemBlue file system, which adjusts replication based on devices~\cite{ensemblue}. Current data replication systems require the user to specify which data should be replicated to which device, and this is often done in a very coarse way (i.e. by file or directory hierarchy). However, being implemented in the OS, personal computing images must handle every user interaction, and can take advantage of this by learning from user behavior. Certain devices are optimal for accessing certain types of data. For example, portable devices such as a music player or a smartphone are better for listening to music while a desktop computer with large screen and a mouse is optimal for modifying a large spreadsheet. While users have lots of data and many devices, the access graph between devices and data will not be strongly connected, giving opportunities to reduce the cost of replication.

Even if a piece of data is available at every device where the user might want to access it, the user also needs an application with which to access the data. As a result, users must manually install the appropriate applications on each device that they may need them on. However, after installation, applications do not remain static -- they have software updates and configuration changes causing them to mutate over time like user data. Personal computing images aim to enable the user to perform the software update or configuration change once on any one of their devices and automatically have that change propagated to all devices. We take an extreme view of what application state should be propagated and consider application memory and execution context as well as on-disk application data as state that the user will want to make visible on all devices. Consider the case where a user has opened a set of files in a spreadsheet application and configured a set of views to perform some analysis of data. Currently, if the user opens the application on another device, she will have to open all the files again and manually put the application in the same state. We believe to truly implement the concept of a single personal computing image, the dynamic state as well as the persistent state of applications must also be simultaneously visible to all devices. When the user starts the application on another device, it should detect that an existing instance is already running, and unless the user specifies they want to start a new instance, automatically migrate the dynamic state of the running instance to the new device. Unlike user data, application state is more difficult to replicate because it is difficult to separate application state from the underlying OS state. For example, an application may have state stored on the file system, in kernel structures, and in shared repositories like the Windows registry. Personal computing images can sidestep such issues by replicating both OS and application state as necessary, giving every device an identical view of the OS.

Finally, responsible users ensure that their data is periodically backed up for durability. As the number of devices increases, users must then backup more and more of their data on these devices. By having a single personal computing image, users must only backup that single image, meaning that the task of backup does not increase with the number of devices. With a personal computing image, state will likely be replicated across devices just for availability. As a result, some amount of durability against device failure is automatically gained just by eagerly replicating OS state across devices. In addition, since the system actively replicates system state continuously, the user will be able to recover state that is much more current than what would be available with periodic backups. Finally, recovery from a device failure is simplified greatly. When the user acquires a new device or replaces a failed one, rather than having to manually install applications and copy data over, she simply adds the personal computing image to it and all her applications and data become immediately available on the new device.

%% file: evolution.tex
\section{Evolutionary Steps}\label{sec:evolution}

Instead of moving directly to a single personal computing image, systems will evolve towards them. In fact, some of the steps in the evolution from independent personal devices towards a unified personal image have already been taken. We view personal computing images as the final step in a series of evolutionary steps that various systems supporting mobility have been working towards.

\paragraph{\textbf{Step 1: Faking it -- Thin Clients:}} Various technologies such as VNC, RDP and THINC~\cite{thinc:Baratto:2005:TVD:1095810.1095837} enable users to access a system from another device. A single system image could be supported by having only a single device run all applications and host all data while the other devices access the single device as thin clients. Naturally, there are many disadvantages to this architecture. In particular, reliable, high bandwidth and low latency network connections must exist for all thin client devices. Second, such a system does not naturally provide any form of durability -- the single server must itself be replicated and backed up for durability and availability. Nevertheless, we view this as the first, natural step in the evolution towards personal computing images.

\paragraph{\textbf{Step 2: Hacking it -- Virtual Machines:}} Client virtual machines that replicate system images for mobility~\cite{collective:Chandra:2005:CCS:1251203.1251222, isr:Kozuch:2002:IS:832315.837557}, are the next natural step. By moving state and execution between devices, users are less dependent on the quality of the network connection -- after the migration is done, execution proceeds locally. However, such systems still have several restrictions that prevent them from fulfilling all the requirements of personal computing images. First, there cannot be any concurrency across devices -- execution can only occur on one device at a time. However, there are instances where users may use different applications simultaneously on different devices, such as when they listen to music on their phone while composing an e-mail on their laptop. Because the unit of execution is a virtual machine, it is not possible to tease the states of these two executions apart to enable simultaneous execution on two devices. Second, a virtual machine-based system must transfer the entire system image to whatever device the user is using, resulting in a significant delay when switching devices. Previous systems either targeted replication across devices that were far apart so that the time the user takes to move from one device to another could be used to hide the transfer time of the VM~\cite{collective:Chandra:2005:CCS:1251203.1251222, isr:Kozuch:2002:IS:832315.837557}, or they require the user to carry parts of the system image virtual machine with them on a USB drive~\cite{moka5:desktop-management}. However, we envision personal computing images to be used on all of a user's devices so they may switch from the desktop machine they are using to a laptop in their bag or a smartphone in their pocket in a matter of seconds.

We believe that systems are currently not yet out of this phase of evolution and there is interesting work yet to be done in this area. One reason this stage of evolution cannot simply be skipped is that virtual machines are an ideal stopgap for fixing problems with legacy OSs without having to discard the wealth of applications written for those applications~\cite{disco}. Without a good number of mature applications, one cannot be sure that the usage model we envision is the correct one. Also, one cannot get a realistic feel for the amounts of data and state that need to be transferred on a device switch. As a result, we are currently working on a system called UnityVM, that will try to overcome some of these challenges of implementing a personal computing image with virtual machines. UnityVM will utilize a distributed hypervisor to move a single personal computing image implemented as a VM across devices, with the goals of supporting fast migration times, durability and low management cost.

\paragraph{\textbf{Step 3: The Real Thing -- The UnityOS, a Distributed OS:}} The final step in this evolution will be implementation of personal computing images as a single distributed OS running on all devices. Our goals are to eventually transform UnityVM into a native OS that does not require a legacy OS as a component. At its core, we see UnityOS as a distributed OS, much like those designed to run on clusters and NUMA multiprocessors. As such, because accesses to remote nodes are expensive, the OS tries to maintain locality between a process and the data it manipulates to minimize such latencies. However, to the administrator, the nodes appear to function as a whole, and users and application developers are largely not aware that they are actually using a distributed system unless they choose to write a distributed application.

UnityOS brings new challenges due to the heterogeneity of the hardware devices. Not only do the devices differ in computational power, storage and network connectivity, but they also have different user interfaces and sensors attached to them. UnityOS needs to be aware that it is not useful to replicate a camera application onto a device that does not have a camera. Likewise, unlike in a cluster where a single user interface exists for all the nodes, the individual nodes in UnityOS will have different user interfaces, requiring UnityOS to adapt its output to whichever device the user is using. Finally, we expect node failure rates to be higher than in previous systems. Even if nodes do not fail unexpectedly, users generally replace their devices every 2-5 years so devices churn when many devices per a user will be common. Thus, UnityOS must be able to support durability of state in the face of this. We elaborate on what we think will be the key challenges to UnityOS in the next section.

%% file: challenges.tex
\section{Challenges}\label{sec:challenges}

UnityOS will face several key challenges, which we elaborate on here and give possible directions for solving.

\paragraph{\textbf{UI Agility:}} Because each device acts as a portal through which the user may access the UnityOS system image, UnityOS must be able to simultaneously support different user interfaces (UI) to the system state. Moreover, UnityOS must be able to adapt UI presentation and controls to the capabilities of each device, whether the device has a large or small screen, a keyboard or a touch-based interface. To do this, the UI of the OS must be composed entirely of soft state, so that if the user switches from one device to another, the UI state can be destroyed and immediately recreated on the new device. This motivates a partitioned approach in OS design, where abstractions separate the core OS from device and hardware-specific aspects. We believe that lessons learned from model-view-controller architectures~\cite{mvc:krasner_description_1988}, as well as hotplug-able architectures~\cite{hotplugable:Kooburat:2011:BBW:1991596.1991602} will point to ways where one can abstract hardware and UI from the core OS so that these may be dynamically changed during execution.

\paragraph{\textbf{Coordination:}} While all the devices contain the same OS image, a directory is needed for any particular device to find the location of an object it does not store locally. In addition, a device must also be able to discover if an object it is storing locally has been modified by another device also storing a replica of that object. Finally, for durability, devices must be able to determine the number of replicas for each object so that each object has an adequate level of replication. To do this, one could implement a fully distributed directory, such as those used in distributed shared-memory processors. Distribution improves performance by amortizing directory requests across all nodes in the system, but complicates the implementation and enforcement of consistency in the system. In addition, it is less resilient to failures since the loss of any node means that the directory will be incomplete. While we believe that users will have many devices, it is likely that a user will only be able to use a small number of them simultaneously. As a result, we believe that it is not necessary to distribute the directory for performance. Therefore, UnityOS should opt for a centralized directory with a small level of redundancy to ensure high availability. To keep load on the centralized directory light, we take inspiration from GFS~\cite{gfs:Ghemawat:2003:GFS:945445.945450}, where the master simply stores the location of blocks, leaving the decision to replicate and the actual replication and storage of blocks to the peers. A similar such master could be one of the devices or a small set of them, or even provided externally by a third party as a management service to the user.

\paragraph{\textbf{Fast Switching:}} Unlike previous systems supporting mobility, we envision that in the majority of the time, users will be switching devices that are located close together. As a result, any state that was modified on the old device and will be accessed on the new device must be transferred in a matter of seconds. Fortunately, we observe that devices located near each other are also likely to have reliable, high quality network connectivity to each other. For example, a laptop and a smartphone the user has next to her desktop in her office are all likely to be on the same wifi network. So, she will have the enough bandwidth to transfer presentation slides she was working on to the laptop or the smartphone she will take with her in the time it takes for her to walk out the door. On the other hand, when the user leaves her office to go home, then UnityOS will have much longer time to transfer the necessary state over the low-quality WAN connection to their home. To do this, UnityOS not only must learn to detect or even predict the user’s behavior, but it must also know the location of all devices relative to each other -- devices that are located close to the one the user is using are more likely targets of a device switch than devices that are far away. One possibility is to take the advantage of the growing number of location-based services available -- today any device with a GPS or even just a Wifi card is capable of determining its location. Another is for devices to detect if they are on the same subnet or network as the device the user is currently using.

\paragraph{\textbf{Architectural Heterogeneity:}} Currently, two instruction set architectures (ISA) dominate personal computing devices, ARM and x86. Applications and code that run on one ISA cannot transparently run on the other. This challenge might be addressed in several ways. First, application binaries can be enhanced to contain two versions of the binary, one for each ISA. While this would allow an application to run on both architectures, it does not intrinsically allow an already running application to be migrated over to the other architecture -- issues such as byte ordering and memory address space layout may make this difficult. An alternative solution is to produce efficient emulators or binary translators that could handle static binaries or possibly even dynamically transferred state. Finally, we concede that this may not be a problem at all if economic forces result in only one dominant ISA for personal devices in the long-term future.

%% file: conclusion.tex
\section{Conclusion}\label{sec:conclusion}

A period of explosive growth in the number of personal devices is now upon us. If the devices continue to operate as separate images of computation, all managed by the user, then the effort required to manage the devices will scale with number of devices. Instead, a single personal computing image, distributed across a user’s devices will decouple the management burden from the number of devices, thus emancipating the user from the ``tyranny of devices." We observe that many of the underlying mechanisms to support personal computing images are already well known, but have been built with a goal and the set of assumptions that are different from UnityOS's. The personal devices of the future will not lack in computational power or resources, but will be inevitably getting lesser and lesser management attention from their owner. Thus, we take the position that we must work to build a distributed OS for personal devices, with the goal of reducing the management burden the user must bear to maintain these devices.